\documentclass[superscriptaddress, 11pt]{revtex4-1}

\usepackage{amsmath}
\usepackage{graphicx}
\usepackage{soul}
\usepackage{color}

\usepackage{lineno}

\begin{document}
\title{Resonance and Damping in Drop-Cantilever Interactions} 

\author{Crystal Fowler}
 \affiliation{
Department of Biological and Environmental Engineering, Cornell University, Ithaca, NY 14853, United States
}
\author{Rehan Marshall}
 \affiliation{
Department of Biological and Environmental Engineering, Cornell University, Ithaca, NY 14853, United States
}

\author{Maeji Son}
\affiliation{
North London Collegiate School, 109708, Singapore 
}

\author{Sunghwan Jung}
\thanks{sj737@cornell.edu}
 \affiliation{
Department of Biological and Environmental Engineering, Cornell University, Ithaca, NY 14853, United States
}

\begin{abstract}
In this study, we investigated the dynamics of a droplet impacting and oscillating a polycarbonate cantilever beam of nine varying lengths. We analyzed the cantilever’s damping and vibration frequency in relation to a resonance length, where the frequencies of the droplet and the cantilever are equal. 
In the pre-resonance length, the beam vibrates at a frequency higher than that of the droplet. Upon reaching resonance, the frequencies of both the droplet and the cantilever align, and the cantilever is out of phase with the oscillation of the droplet’s apex. This leads to increased damping rates. At this resonance length, the droplet’s force and the direction of the cantilever oppose each other. When the cantilever length exceeds the resonance length, it synchronize more with the droplet apex. This alignment allows the droplet force and the cantilever to work in phase. Our findings provide fundamental insights into the damping effect of droplet impacts on elastic surfaces around resonance.

\end{abstract}

\maketitle
\section{Introduction}

The dynamics involved in the interaction of droplets and elastic surfaces are relevant to many biological and industrial applications. The biological applications include understanding the water retention from the canopy of the rain forest. Water interaction with the canopy is important to modeling the forest ecosystem \cite{dawson2018value}. 
Plant leaves have many biological factors that can alter surface characteristics (i.e. hydrophobicity \cite{holder2013effects}, shape, and surface topography \cite{kang2018seasonal,ecological:perspective}), which affect the canopy water storage capacity \cite{holder2013effects}. Furthermore, the geometry and material properties of the leaf can differ across species and position in the canopy \cite{ecological:perspective}. 
Bhosale et al. have used droplets of various sizes and real leaves to mimic the leaf-raindrop system, which allows proper insight into droplet impact location and torsion measurements \cite{bhosale2020bending}. Other studies have simplified the leaf-raindrop system with a droplet and cantilever system; therefore varying only surface hydrophobicity, cantilever material properties, droplet diameter and droplet velocity \cite{nanko2016differences}. Modeling with a droplet and cantilever system also has helped understand the distribution how spores are spread \cite{kim2019vortex,wu2024coherent}. 

Examples of droplet-cantilever applications include piezoelectric energy harvesting with rain drops \cite{leaf:drop:PhysRevApplied,wong2015harvesting} and application of pesticide sprays on leaves \cite{smith2000droplet}. As such, considerable attention has been posed to technological studies in applying biomimetics of raindrop oscillation to create self-sustained systems for industrial applications. For instance, the droplet impact could be used to harvest vibration energy from piezoelectric material. The rain drop harvesting model has been developed by Guigon et al. and is now further developed to link bridge structures for high rate of conversion of high voltages \cite{guigon2008harvesting}. In recent studies, industrial usage has further reached the extent of developing an energy-harvesting device that is embedded as an artificial elastic plant leaf and creates electricity from raindrops \cite{ilyas2015piezoelectric,leaf:drop:PhysRevApplied}. 
Other work on the droplet and leaf model have made a full theoretical model between the leaf and the droplet \cite{lauderbaugh2022biomechanics}. Some theoretically determine the droplet trajectory after it impacts a super-hydrophobic cantilever \cite{fang2023target}. While many have made the investigation into the leaf-droplet interaction, there still exists a need to investigate the damping of the cantilever and the dynamics between characterizing the droplet and the surface. 

For droplet impact on a rigid surface, many studies have quantified the droplet-substrate relationship through contact time, contact lines, spreading velocity, and spreading height. It has been found that higher frequencies of the droplet are apparent in the mobile contact line of the first few oscillations when compared to pinned contact lines on a hydrophilic surface \cite{contactline:droplet}. A similar study that focused on droplet with a range of hydrophilic to hydrophobic substrates found that higher-order modes and pinned modes decay faster \cite{Kern:Bostwick:Steen:2021}. 
Depending on the surface tension, the frequency and mode shape of the droplet have been verified experimentally \cite{periodic:droplet:table}. On a vibrating surface, a droplet can display different morphology based on its resonance modes \cite{sharp2012resonant}.
In addition, the resonance of the droplet in a vertical direction on a hydrophobic surface was experimentally and theoretically determined by Shin et al. by investigating the pressure difference of the droplet surface \cite{shin2014shape}. At these resonance modes, the droplet demonstrated 'jumping up out of the surface', which resulted in greater droplet expansion, contraction, and jumping height. All of these studies have demonstrated the characteristics and complex dynamics that a droplet can exert on a rigid substrate, however applications of these theories demand to have external vibrations as a factor. Thus, these external vibrations of the substrate can be modeled with a droplet impact on a cantilever. 

In a cantilever-droplet coupled system, the interplay of various forces between the droplet and the cantilever varies depending on both the length scale and time scale. For instance, Liu et al. found that a smaller size of the cantilever and droplet can create more significance of capillary forces on the cantilever \cite{micro-cantilever:droplet}. Furthermore, Weisensee et al. found that the contact time and the resulting droplet dynamics are dependent on the impact conditions of the droplet \cite{PhysRevFluids:superhydrophobic}. In a similar study, Upadhyay et al. modeled with a spring and mass system the contact time between the superhydrophobic cantilever and the droplet \cite{upadhyay2021bouncing}. Both of these studies only investigated the instant impact that the droplet had on the cantilever and not resulting droplet-cantilever interactions. However, Kumar et al. analyzed the remaining oscillations after the droplet impact on a hydrophobic cantilever in their study \cite{coupled:dynamics}, including the damping rate of the cantilever and and an energy component analysis. Based on a previous study, Huang et al. found an inverse relationship between the contact angle of the droplet and the initial deflection of the cantilever \cite{huang2019droplet}. This information is also supported by Gart et al. that found a higher torque value of a hydrophilic cantilever versus a hydrophobic cantilever \cite{leaf:drop:PhysRevApplied}. Thus, it would be beneficial to include a more hydrophilic surface to obtain more deflection in the cantilever; therefore changing the subsequent damping ratios. The cantilever, instead of metal and hydrophobic, should possibly be a polymer cantilever that will result in more applications in the development of rain drop harvesting \cite{free:cantilever:beams}. 

The assumptions made for its experiment are such as (1) no external force applied (except for mass as a restoring force and a frictional force) and (2) points of inflection occur at mid span of cantilevers. These were designed in such a way as the vibration frequency ($\omega$), and damping ratio ($\zeta$) must be shown clearly. Under the following setting, the vibration will adapt the features of the harmonic oscillator, where the dissipation of energy created due to the non-conservative forces and its critical damping returns the system to equilibrium over time \cite{strang:textbook}. Through the damped harmonic oscillator, the damping ratio ($\zeta$) could be applied to access the behavior of systems which could be overdamped ($\zeta > 1$), underdamped($\zeta < 1$) or critically damped ($\zeta = 1$) \cite{strang:textbook}. 

Since the fundamentals of these applications lies in the leaf-drop dynamics, there is a need for in-depth research on understanding how the damping effect of droplets impact on hydrophilic surfaces. This present work is to understand the droplet-cantilever dynamics by measuring the frequency, cantilever damping, droplet modes, phase shift, and vertical displacement of the majority of oscillations present.


\section{Methods and Materials}
\vspace{-5mm}
\subsection{Experimental Setup}
Our experimental setup consists of two main components, a drop dispenser and a poly-carbonate cantilever. For the drop dispenser, a syringe pump (New Era Pump Systems Co.) was faceted with a needle of 2 mm in diameter, thus resulting in droplets of approximately 3.8 mm in diameter. The diameter of the droplet was averaged over five trials with images taken before impact on the cantilever. The mass of the droplet was calculated with $\rho$ as density and $d_o$ as initial droplet diameter in the equation $m_{drop} = (1/6)\pi \, \rho \,d_{o}^{3}$. The needle was positioned 150 mm above and orthogonal to the cantilever, releasing droplets with an impact speed of approximately 1.43 m/s. The impact speed $V$ was calculated with gravity $g$ and height of syringe $h$ in $V=\sqrt{2gh}$. The corresponding Weber number ($We = \frac{\rho V^{2}d_{syringe}}{\gamma_{LG}}$) is 46-68, where $d_{syringe}$ is the syringe diameter and $\gamma_{LG}$ is the surface tension of water ($7.2 \times 10^{-2}\: \rm N/m$). Deionized water was used to produce the droplets. 

\begin{figure*}[hbt!]
\centering
\includegraphics[width=1\textwidth]{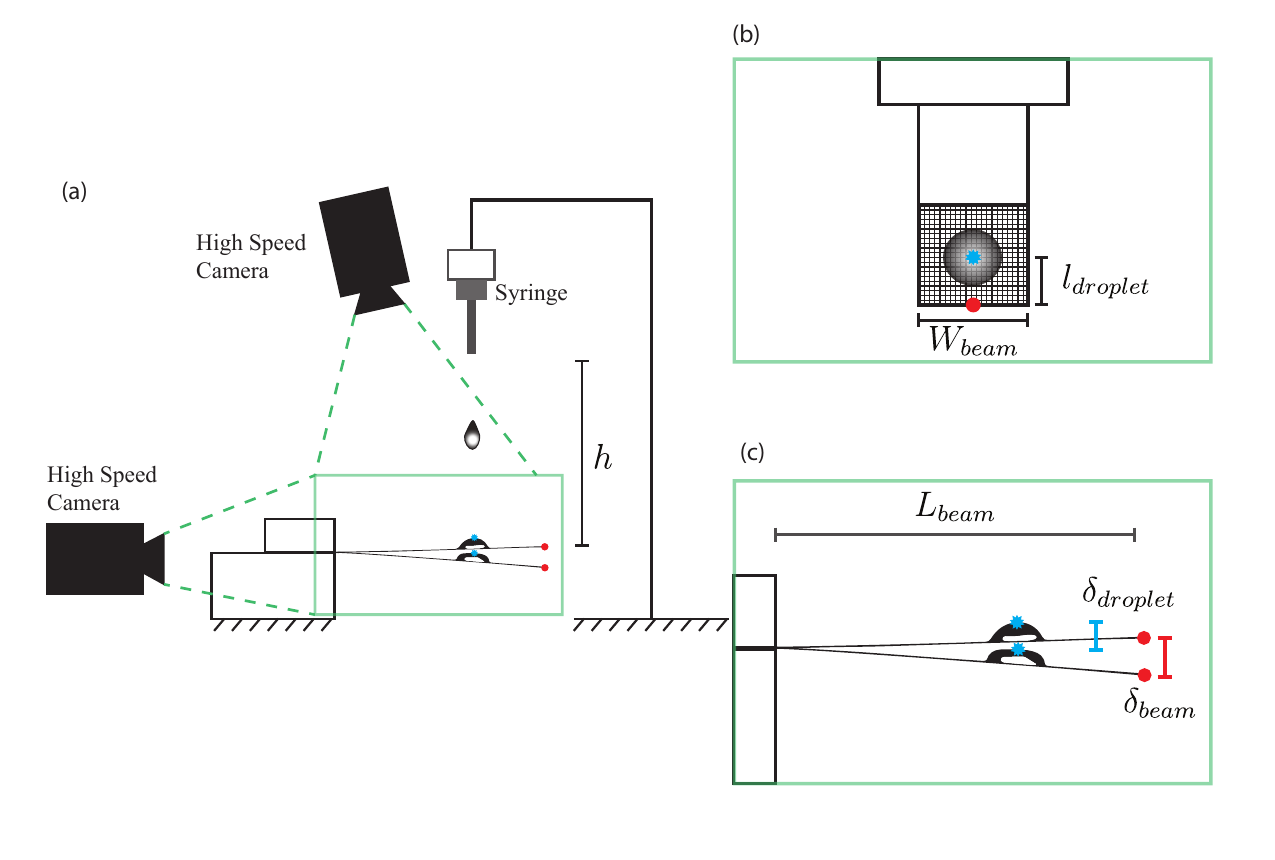} 
\caption{ (a) The cantilever was held in place by b) the linear stage and clamp in front of the c) background light source. (d) 15 cm above the cantilever the syringe was connected to a pump. The syringe was positioned to drop the droplets 1 cm from the edge of the cantilever. }
\label{setup}
\end{figure*}

\begin{table*}[hbt!]
\begin{tabular}{|l|l||l|l|}
\hline
Droplet Diameter      & 3.8\:mm                     & Cantilever EI                                & $6.77\times10^{-5}\: \rm Nm^{2}$  \\ \hline
Droplet Impact Speed  & 1.43\:m/s                   &  Cantilever Density                          & $1.17\times10^{3}\: \rm kg/m^{3}$ \\ \hline
Weber Number          & 46-68                          & Cantilever Thickness                         & $0.254\: \rm mm$\\    \hline  
Pumping Rate          & $0.3\: \rm mL/min$          &  Cantilever Width                            & $20\: \rm mm$\\  \hline 
Surface Tension       & $7.2\times10^{-2} \: \rm N/m$  &  Cantilever Young's Modulus \cite{mat:cite}  & $2.41\: \rm kg/m^{3}$\\ \hline
Static Contact Angle  & $82^{\circ}$                & Cantilever Poisson's Ratio \cite{mat:cite}   & 0.37\\ \hline   
Needle Diameter       & 2\:mm                       &  Needle Height                               & 150\:mm \\ \hline
\end{tabular}
\caption{The physical values of the droplet, cantilever, and needle involved in the experimental setup. }
\label{table1}
\end{table*}

For our experiments, we used poly-carbonate cantilevers, which was cut with the same length and width using a silhouette cutter (Silhouette America Portrait Co.). Another cantilever was cut from the poly-carbonate sheet that was printed with a grid pattern (Figure \ref{setup}). 
Cantilevers were rinsed with DI water first, then rinsed with isopropyl alcohol and allowed to dry. The cantilever lengths were attained by clamping the 80 mm cantilever at the desired length. The cantilevers were clamped by resting them on top of a linear stage and using a screwed in clamp to adjust the length of the cantilever. The cantilevers were positioned such that they were orthogonal to the drop-dispensing needle and the droplet would impact approximately 10 mm from the free end of the cantilever to minimize droplet spillage. The length of these cantilevers was varied from 20-50 mm. To capture the droplet-cantilever interaction, side and top views of the apparatus were captured using high-speed cameras. The high-speed cameras (Photron FASTCAM NOVA S9) were positioned above and on the side of the apparatus as shown in Figure \ref{setup}c and Figure \ref{setup}b. The camera settings for the high speed cameras were 750 frames per second with 4000 per second shutter speed. The frame resolution was 1024 pixels by 1024 pixels. 

\subsection{Cantilever Rigidity}
The flexible rigidity depends on the the elastic modulus $E$ and the second moment of area $I$. The second moment of area $I$ comprises of the width $b$ and thickness $h$ of the cantilever, where $I=\frac{bh^{3}}{12}$. The $EI$ values of these cantilevers were experimentally calculated for the normal and grid cantilevers respectively (Table \ref{table1}). To calculate the $EI$ values, we measured the deflection of both the grid and normal cantilevers using a force sensor (Futek Miniature S-Beam Jr. Load Cell). We recorded the cantilever maximum tip deflection using a DSLR camera (Nikon DSLR D7100). Dependent on the cantilever length $L$, the force $F$ to deflection $\delta$ ratio is obtained by a singular force point on a cantilever cantilever: $F=\delta\frac{3EI}{L^{3}}$. Graphing the force from the force sensor versus the deflection of the cantilever from the camera yields the slope (${3EI}/{L^{3}}$). Then, the $EI$ value is calculated by rearranging the equation. A total of ten trials were done to get a good estimation of the flexible rigidity of the cantilever. 

\subsection{Video Analysis}
To analyze the free vibration patterns of both the cantilever and droplet, we first measured the tip of the cantilever using Tracker (Open Source Physics). The sides of the droplet was tracked with a MATLAB-based DLTdv8a \cite{hedrick2023tracking}. We tracked two points adjacent to the contact point of the cantilever and droplet. Then with our custom MATLAB script, the outline of the droplet was extracted within the range of the two contact points of the droplet edges to the cantilever. The midpoint of the two droplet edges was used to approximate the droplet apex. There was ten trials done for every cantilever length to get a good representative of the droplet-cantilever behavior. Example oscillations of the different cantilever lengths are in Appendix section B.

\section{Results}
\subsection{Cantilever length comparative analysis} 

\begin{figure*}[hbt!]
\centering
\includegraphics[width=1\textwidth]{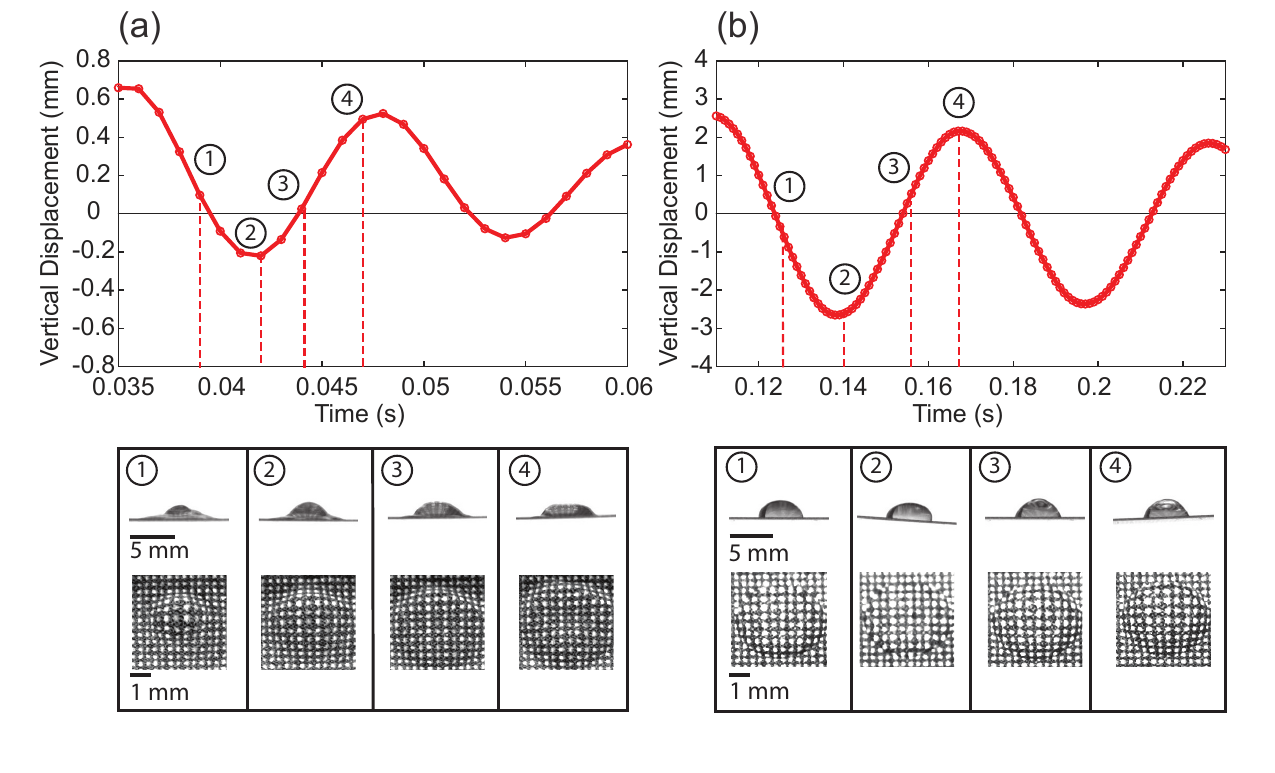}
\caption{ (a) The third oscillation of the cantilever tip displacement for 20 mm length cantilever. At times 1, 2, 3, and 4, the the side view and the top view of the cantilevers with the droplet can be seen. (b) The third oscillation of the droplet and cantilever tip displacement for 50 mm length cantilever. The 50 mm cantilever length at positions in 1, 2, 3, and 4 are over a single period of the cantilever oscillation.}
\label{fig1}
\end{figure*}

Figure \ref{fig1}a shows uniform oscillations for the 20 mm cantilever length. The crest of the cantilever tip at time ‘4’ does not align with the highest height of the droplet apex observed at time ‘2’. Instead, the droplet exhibits a concave depression at its center at time ‘4’ as shown in the lower planel. In contrast, for the longer cantilever in Figure \ref{fig1}b, the droplet does not deform much while the beam vibrates.

\begin{figure*}[hbt!]
\centering
\includegraphics[width=1\textwidth]{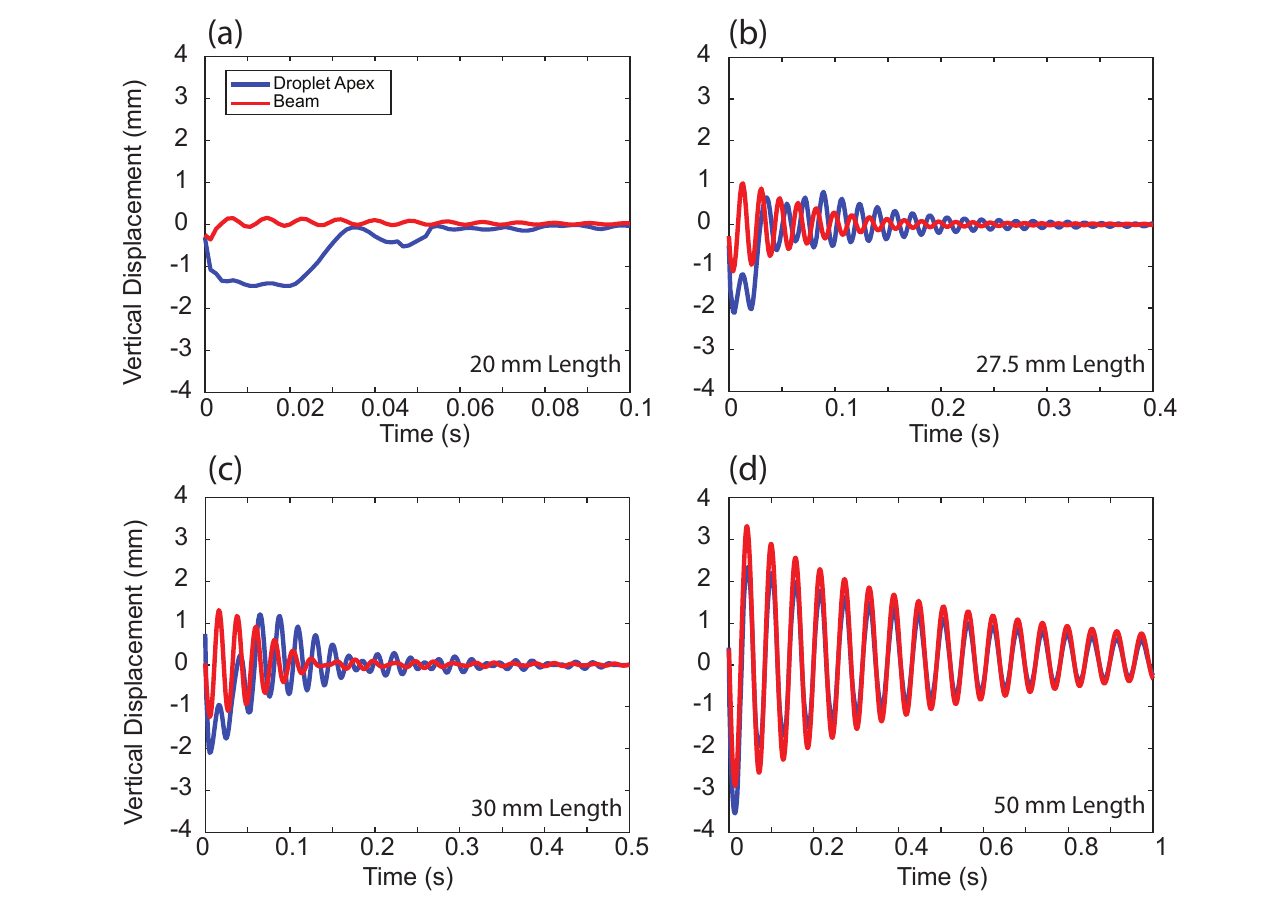}
\caption{(a) Droplet apex and cantilever tip displacement for 20 mm length cantilever. The cantilever shows uniform oscillations, but the droplet apex behaves like the short cantilever is a rigid surface. (b) Droplet apex and cantilever tip displacement for the 27.5 mm length cantilever, in which the droplet frequency matches the natural frequency of the cantilever. In addition, the cantilever and droplet apex are mostly out of phase. (c) Droplet apex and cantilever tip displacement for 30 mm cantilever length shows a second mode characteristic in which there are two frequencies being displaced in both the droplet and tip. (d) Droplet apex and cantilever tip displacement for the 50 mm cantilever length shows that both the droplet and the cantilever have similar frequencies and are in phase. }
\label{fig2}
\end{figure*}

There are significant differences in the amplitude and frequency between the droplet and the cantilever as the cantilever length varies. The vertical displacement was measured from a position when the cantilever is at equilibrium. For a cantilever of 20 mm length, the amplitude of the beam oscillation is less than 1 mm, while the apex of the droplet exhibits slow deformation and negligible oscillation (refer to Figure \ref{fig2}a). In the intermediate lengths (27.5 mm and 30 mm), the first 3-4 oscillations do not follow simple oscillations due to the initial drop impact (Figure \ref{fig2}b and Figure \ref{fig2}c).  

\begin{figure*}[hbt!]
\centering
\includegraphics[width=1\textwidth]{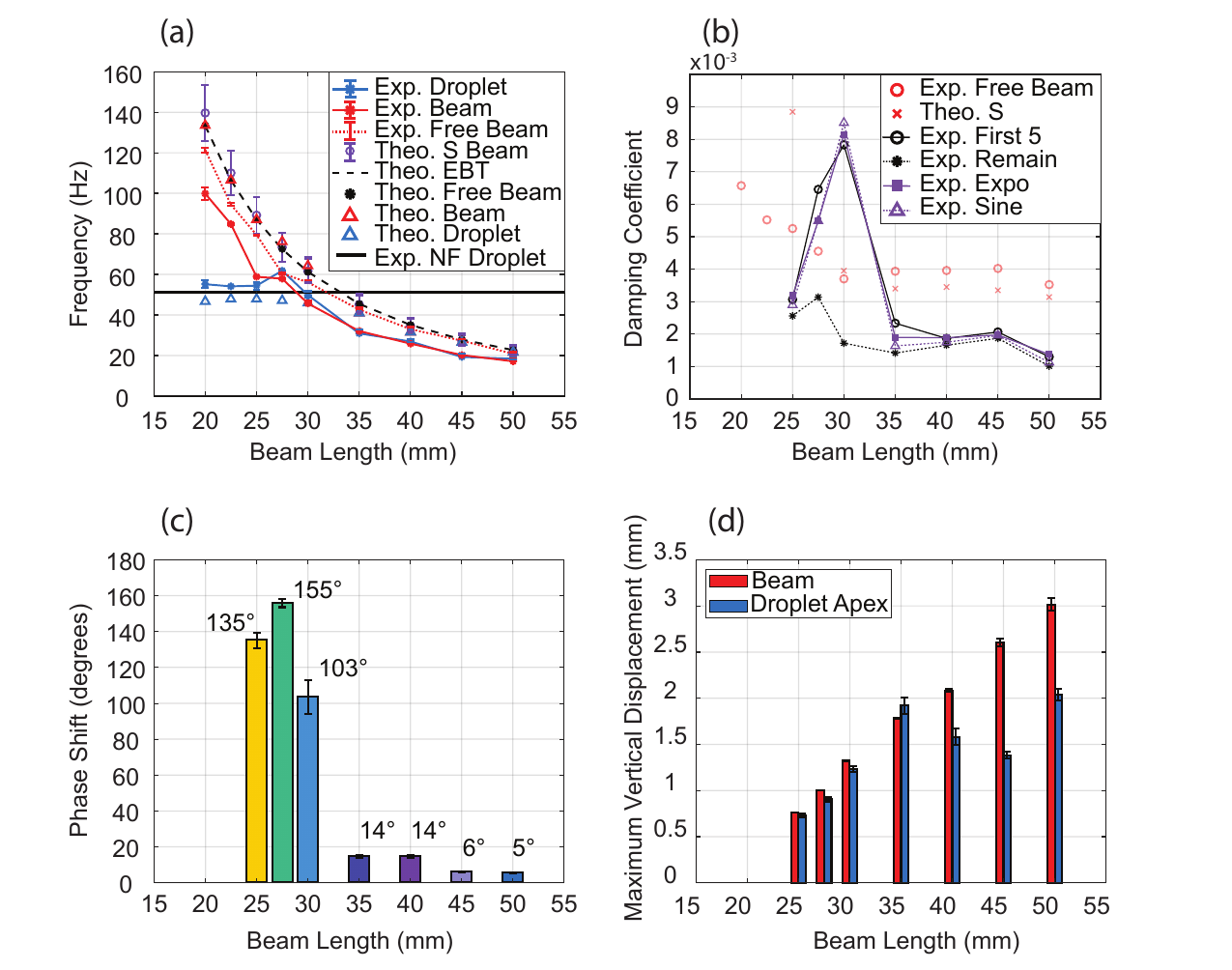}
\caption{ (a) The frequencies of the droplet apex (Exp. Droplet), cantilever tip (Exp. Beam), free vibration cantilever tip (Exp. Free Beam), simulation cantilever (Theo. S Beam), Euler-Bernoulli Theory (Theo. EBT), the cantilever in a single mass and spring system (Theo. Free Beam), the cantilever in the two spring and mass system (Theo. Beam), the droplet in the two spring and mass system (Theo. Droplet), and the natural frequency of the droplet (Exp. NF Droplet) are plotted against cantilever length. (b) The tracking of the cantilever tip for each of the lengths was fit with two equations and compared against a free vibration cantilever (Exp. Free Beam). Fitting all oscillations with a damped sine wave (Exp. Sine) and an exponential decay function (Exp. Expo) was used for the droplet-cantilever system. The first five (Exp. First 5) and remaining (Exp. Remain) oscillation damping were included as well. (c) The phase shift between the positive peaks between the droplet apex and cantilever tip. The 20 mm and 22.5 mm length were not graphed due to the inconsistency of the oscillations. (d) The maximum deflection comparison (highest amplitude) for each cantilever. 
}
\label{fig3}
\end{figure*}

First, we utilized the Fast Fourier Transform (FFT) to measure the frequencies of both the droplet apex and the cantilever tip. The results of these measurements are displayed in Figure \ref{fig3}a. The Euler-Bernoulli Equation (EBT) frequency $f_{beam}$ is based on the flexible rigidity $EI$ of the cantilever and length $L$ as 
\begin{equation}
    f_{beam}\simeq \frac{3.516}{2\pi}\sqrt{\frac{EI}{m_{c} + m_{drop}}} \frac{1}{L^{3/2}} \,.
    \label{EBT}
\end{equation}  
The theoretical EBT frequency predicts slightly higher frequency of the experimental beam oscillation.  
On the other hand, the oscillation frequency of the droplet remains relatively constant up to 30 mm, which is close to the Rayleigh's frequency (see Section III.B). Beyond this point, the droplet’s frequency aligns with that of the beam.
Next, we evaluated the damping coefficient of the beam vibration, as shown in Figure \ref{fig3}b. A higher damping coefficient is indicative of the droplet’s enhanced damping effect on the cantilever, particularly at a resonance length around 30 mm. Generally, we observed a pronounced decay in the beam oscillation during the initial five oscillations, compared to the oscillations that follow. 


We also quantified the phase shift in oscillation, which is determined by the difference between the droplet and beam oscillations. In our observations, the cantilever typically exhibits the leading oscillation, while the droplet apex shows the lagging oscillation.
For the cantilever lengths of 25 mm to 30 mm, a significant phase shift greater than $100^{\circ}$ was observed, which corresponded with higher damping coefficients, as shown in Figure \ref{fig3}c. However, for cantilever lengths exceeding 30 mm, the phase difference between the droplet and the cantilever was minimal. 
The maximum vertical displacements of the droplet and cantilever are shown in Figure \ref{fig3}d. Generally, the maximum vertical displacement of the cantilever increased with cantilever length. However, the maximum vertical displacement for the droplet apex showed an increase only up to the 35 mm cantilever length, and then decreased. 



\subsection{Modeling with a spring and mass system} 
The cantilever system can be modeled with a singular mass $m_{c}$, a spring, and a damper. 
The spring constant of the cantilever depends on its flexual rigidity $EI$ and length $L$ as 
\begin{equation}
    k_{c}=\frac{3EI}{L^{3}},
    \label{beam_k}
\end{equation}  
The equation for a single spring and mass system is given as
\begin{equation}
    \ddot{y} = \frac{1}{m_{c}}[(-c_{c})\dot{y}-k_{c}y],
    \label{single_system}
\end{equation}  
where $c_{c}=2m_{c}\lambda_{c}$. Here, the experimental damping coefficient $\lambda_{c}$ for the cantilever in Figure \ref{fig3}b was used.  

Next, we used a two mass and spring system to analyze the coupling between the droplet and the cantilever. This system included two distinct dampers for the droplet and the cantilever. The damping term for the droplet was measured from experiments conducted on a rigid surface (Appendix section c). The spring constant of the droplet $k_{drop}$ is determined by Raleigh's frequency equation of a freely oscillating droplet. The frequency depends on the surface tension $\gamma_{LG}$, density $\rho$, and initial droplet size $d_{o}$ \cite{rayleigh:droplets}: 
\begin{equation}
f_{drop} \simeq \sqrt{ \frac{\gamma_{LG}}{(\pi/6)\rho d_{o}^{3}} }
\label{drop_f}
\end{equation}
The spring constant for droplet can be estimated from the the Rayleigh frequency as \cite{coupled:dynamics}
\begin{equation}
    k_{drop}={4\pi}^2\gamma_{LG},
    \label{drop_k}
\end{equation}  
Therefore, the cantilever modeled with the two mass and spring system can be characterized by the following equations
\begin{equation}
    \ddot{y_{c}} = \frac{1}{m_{c}+m_{drop}}[-(c_{drop} + c_{c})\dot{y_{c}} + c_{drop}\dot{y}_{drop} - (k_{drop} + k{c})y_{c} + k_{drop}y_{drop}],
    \label{cant_model_eqn}
\end{equation}  
\begin{equation}
    \ddot{y}_{drop} = \frac{1}{m_{drop}}[c_{drop}\dot{y_{c}} - c_{drop}\dot{y}_{drop} + k_{drop}{y_{c}} - k_{drop}y_{drop}],
    \label{drop_model_eqn}
\end{equation}  
The derivation of Eq. \ref{cant_model_eqn} and Eq. \ref{drop_model_eqn} is further explained in Appendix section D. MATLAB's ODE45 solver was used to solve for the displacement of the cantilever and droplet. The frequencies of the droplet (blue triangles) and cantilever (red triangles) were plotted in Figure \ref{fig3}a. 

\subsection{Simulations} 
We used the Structural Model object in MATLAB to simulate the natural frequency response of the cantilever for each length. First, using AUTODESK Fusion 360, a stl file was made with the cantilever's dimensions and imported as a Partial Differential Equation (PDE) toolbox in MATLAB. The PDE model object was initiated with the Young's Modulus, Poisson's Ratio, and mass density of the cantilever (Table \ref{table1}). Due to the material thickness tolerance, for each of the cantilever lengths a $\pm$10\% thickness tolerance was applied to the models and can be seen as error bars in Figure \ref{fig3}a. The frequency values with tolerance ranges (blue circles) are in good agreement with the EBT frequency values but slightly higher than experimental frequencies, as seen in Figure~\ref{fig3}a. 

\begin{figure*}[hbt!]
\centering
\includegraphics[width=0.95\textwidth]{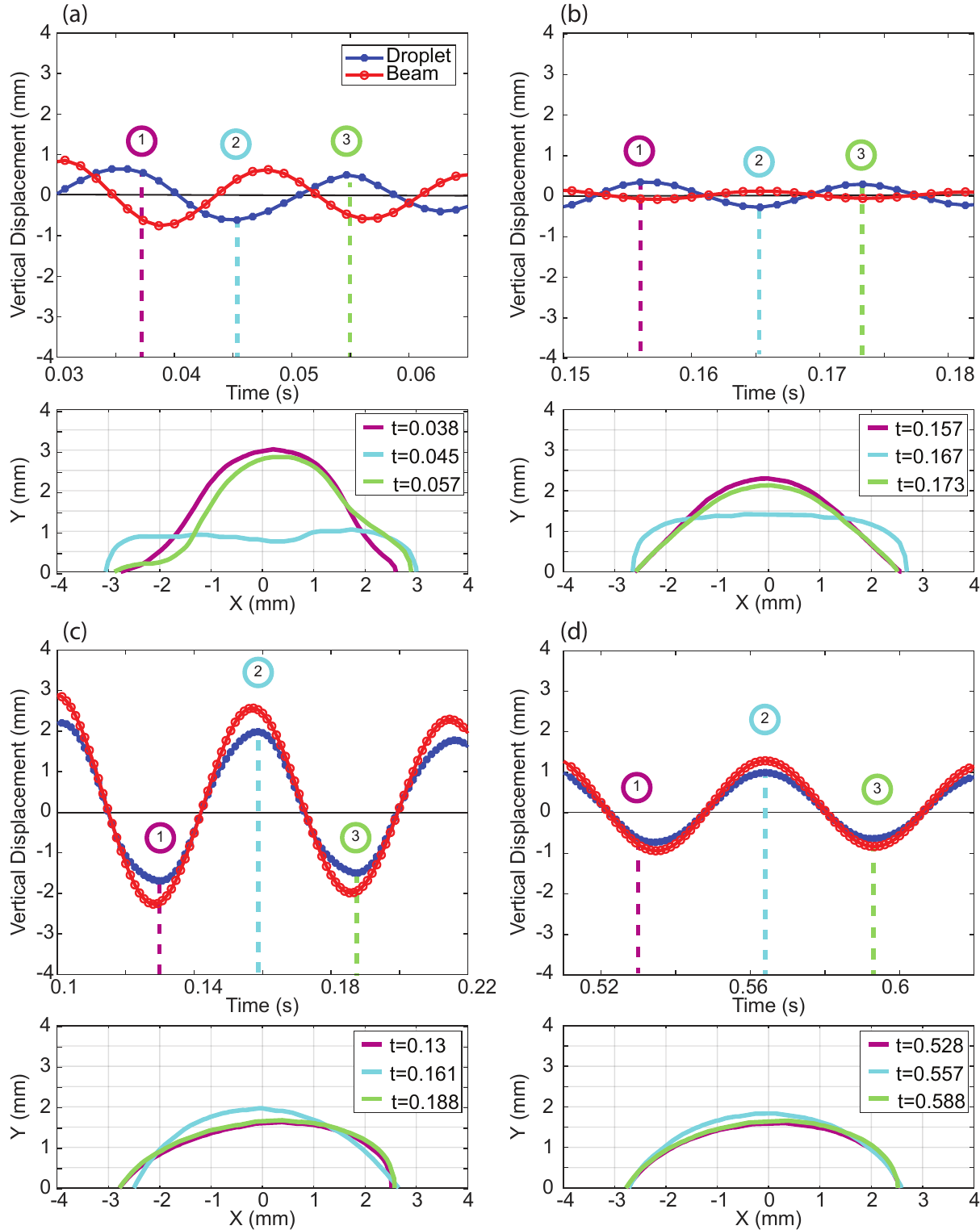} 
\caption{The outline of the droplet and the corresponding time stamps for the (a) third oscillation and the (b) tenth oscillation for the 27.5 mm cantilever length.  The droplet outline and corresponding droplet apex and cantilever tip displacements for the (c) third and (d) tenth oscillation for the 50 mm cantilever length. }
\label{fig4}
\end{figure*}

\subsection{Droplet outline and modes} 

Droplets on vibrating surfaces change the droplet morphology based on the surface frequency \cite{Kern:Bostwick:Steen:2021}. Figure \ref{fig4} showed the morphological change in droplet oscillation at different times over a single oscillation period. The droplet outline was rotated by the slope of the cantilever at that time to make the base of the droplet at 0 degrees. 
Figure \ref{fig4}a and Figure \ref{fig4}b presented the outlines of the droplet at the third and tenth oscillations, respectively, for a cantilever length of 27.5 mm. On the other hand, Figure \ref{fig4}c and Figure \ref{fig4}d displayed the outlines for a cantilever length of 50 mm.  
When the droplet reached its minimum height, the potential energy transitions into interfacial energy and vice versa. The forces at the interface depend on the tensions among the liquid, vapor, and solid. This force balance is also characterized by whether the contact line is mobile or pinned \cite{mondal2023physics}. Previous studies have explored the dynamics of mobile versus pinned contact lines on vibrating surfaces. In our experiments, we observed mobile contact lines during the early oscillation stages and pinned contact lines in the later stages of oscillation.  

\subsection{Cantilever damping and droplet damping influence}
\begin{figure*}[hbt!]
\centering
\includegraphics[width=1\textwidth]{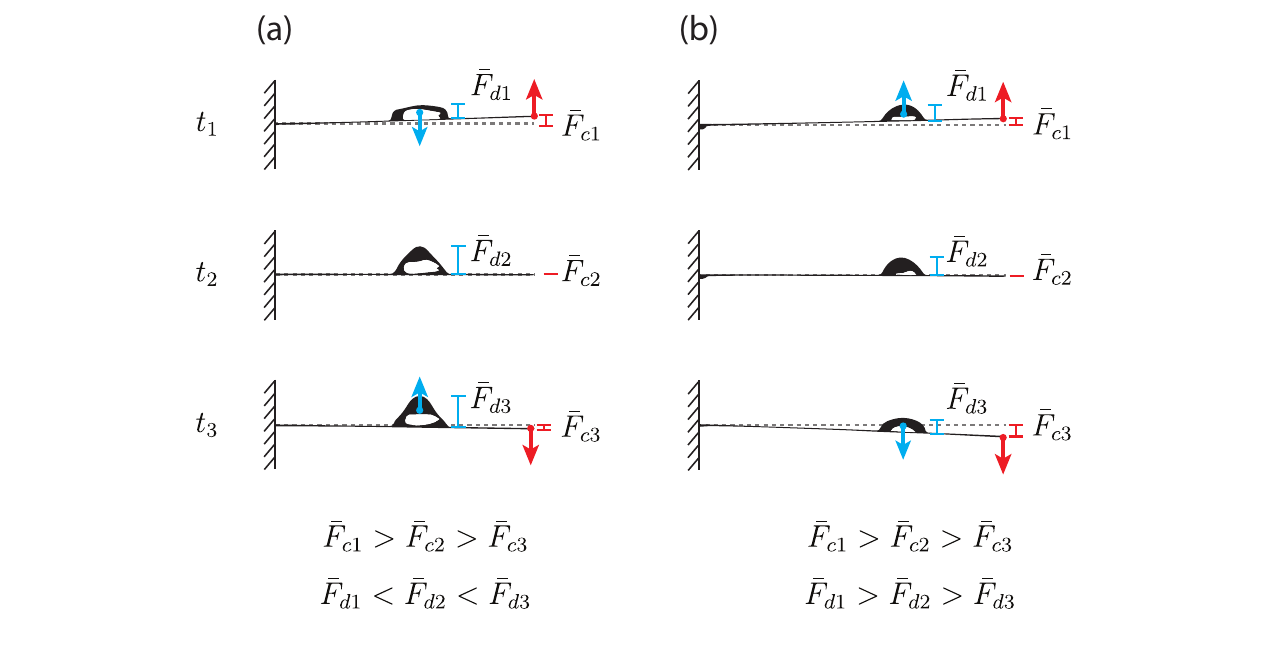} 
\caption{The outline of the droplet and cantilever for an (a) out-of-phase and (b) in-phase  oscillation. }
\label{fig8}
\end{figure*}

Near the resonance length (27.5 to 30 mm), the droplet’s frequency gets close to the cantilever’s frequency and the cantilever oscillates out of phase with the droplet motion. Figure \ref{fig8}a illustrates how the force direction of the droplet ($F_{d}$) moves in the opposite vertical direction of the cantilever force ($F_{c}$), which quickly damps out the cantilever vibration.
For cantilevers longer than the resonance length, the cantilever and droplet are essentially in phase (as shown in Figure \ref{fig8}b), which does not provide additional damping to the cantilever. 

\section{Discussions} 

Our study investigated the interaction between a droplet and a cantilever. Especially, we focused the damping effect of the cantilever in response to the droplet oscillation. We used nine different lengths of a polycarbonate cantilever to characterize the damping in a droplet-cantilever system. Key parameters such as frequency, phase shift, overall maximum amplitude, and the damping coefficient were measured to understand the influence of the droplet interacting with the cantilever. The cantilever frequency decreases with its length as predicted by the Euler Beam Theory (\ref{EBT}), while the drop oscillation frequency remains constant close to the Rayleigh's frequency (\ref{drop_f}). At a resonance cantilever length (around 30 mm), the cantilever frequency aligns with the droplet oscillation frequency. To find the resonance cantilever length theoretically, we equated both frequencies as follows 
\begin{eqnarray}
f_{beam} &=& f_{drop} \\
  \frac{3.516}{2\pi}\sqrt{\frac{EI}{m_{c} + m_{drop}}} \frac{1}{L_{resonance}^{3/2}} &=& \sqrt{\frac{\gamma_{LG}}{(\frac{\pi}{6})\rho d_{o}^{3}}} \\
 L_{resonance} &=& \left( \frac{3.516}{2\pi} \right)^{2/3} \left( {\frac{EI}{m_{c} + m_{drop}} \, \frac{(\frac{\pi}{6})\rho d_{o}^{3}}{\gamma_{LG}} } \right)^{1/3} \,
\end{eqnarray} 
The predicted resonance length is 32.9 mm, which is close to 30 mm observed in experiments. 

In our experiments, we observed that cantilever lengths less than the resonance length (30 mm) demonstrated increased damping, phase shift, and vertical droplet displacement. Conversely, cantilever lengths exceeding the resonance length exhibited reduced damping, in-phase oscillations, and a diminished vertical displacement of the droplet apex. 
When the cantilever length got close to the resonance length, the cantilever displayed higher damping rates during the initial oscillations compared to cantilevers of other lengths. Our findings indicated that these increased damping rates were most pronounced within the first five oscillations of the cantilever.

However, there are limitations in our study. The nature of the droplet’s contact-line, whether mobile or pinned, could significantly influence its damping and vibration. A more detailed examination of the contact-line and the droplet’s vibration mode could enhance our understanding of the higher frequencies observed in the initial oscillations when the contact-line was mobile.
Moreover, the decaying modes of the droplet are presumed to contribute to the cantilever’s damping rate. Our research can also help translate the characteristics of the cantilever into features that can optimize the design of a rain harvester, contribute to ecological modeling, and find utility in other applications.

\section{Appendix} 

\subsection{Droplet and cantilever}
An example of analyzing the frequency, damping coefficient, phase shift, and maximum vertical displacement for one trial. FFT was used to determine the frequency of the droplet apex and cantilever.
\begin{figure*}[hbt!]
\centering
\includegraphics[width=1\textwidth]{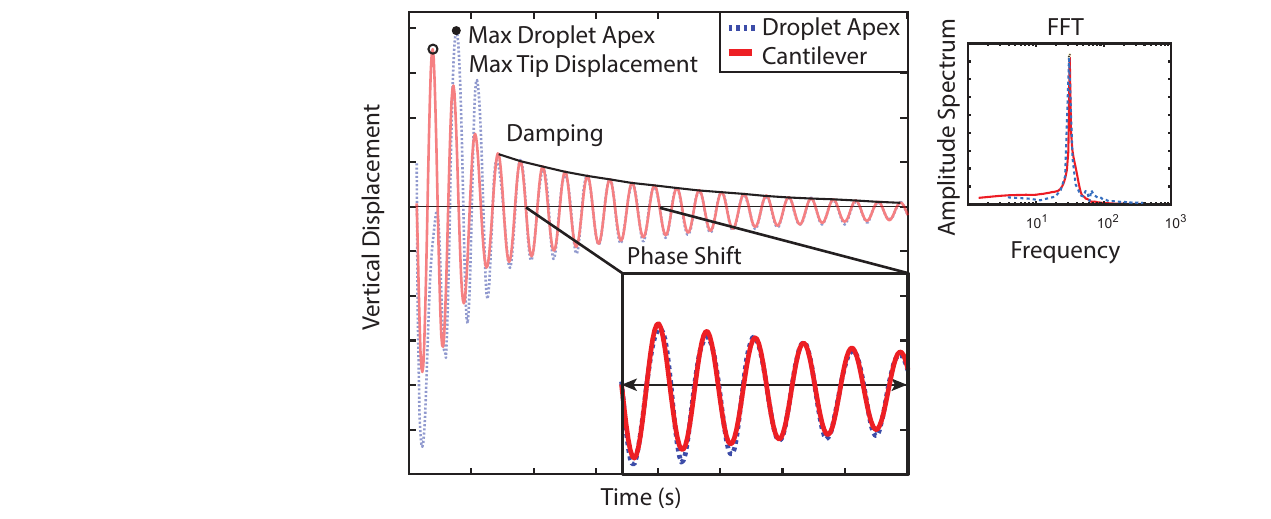} 
\caption{}
\label{fig_S1}
\end{figure*}

\subsection{Example oscillations}
Side profile of the third cantilever oscillation of the droplet and cantilever system for cantilever lengths 20 mm (a), 27.5 mm (b), 30 mm (c), and 50 mm (d).
\begin{figure*}[hbt!]
\centering
\includegraphics[width=0.99\textwidth]{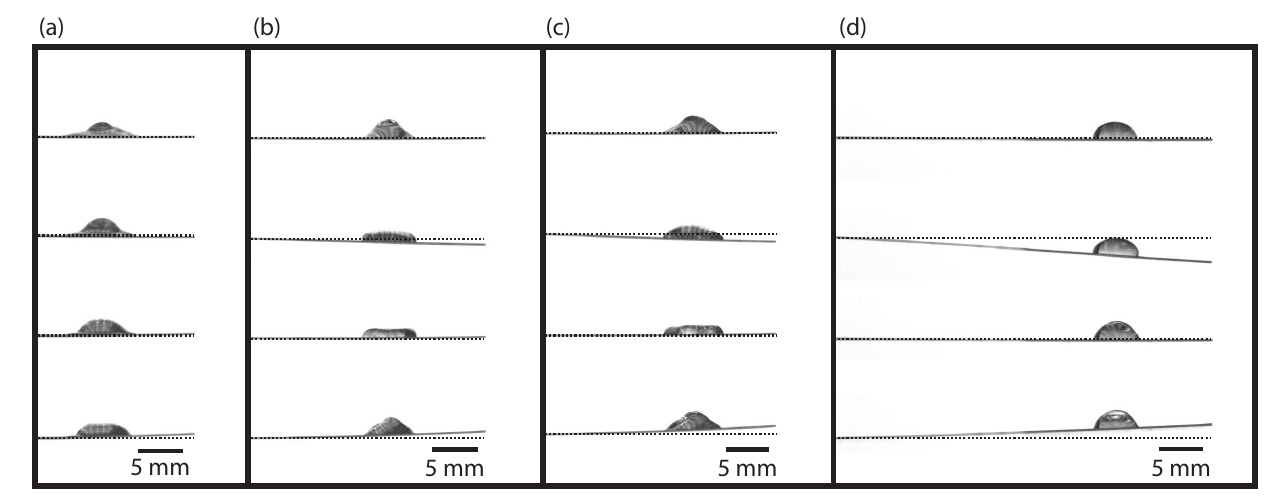} 
\label{}
\end{figure*}

\subsection{Droplet natural frequency}
The natural frequency of a droplet on a rigid hydrophilic surface example trial. A total of five trials were used to compute the mean exponential decay fitting $\lambda_{drop}$. The exponential decay fitting was then used to compute the damping coefficient of the droplet $c_{drop}=2m_{drop}\lambda_{drop}$.
\begin{figure*}[hbt!]
\centering
\includegraphics[width=0.99\textwidth]{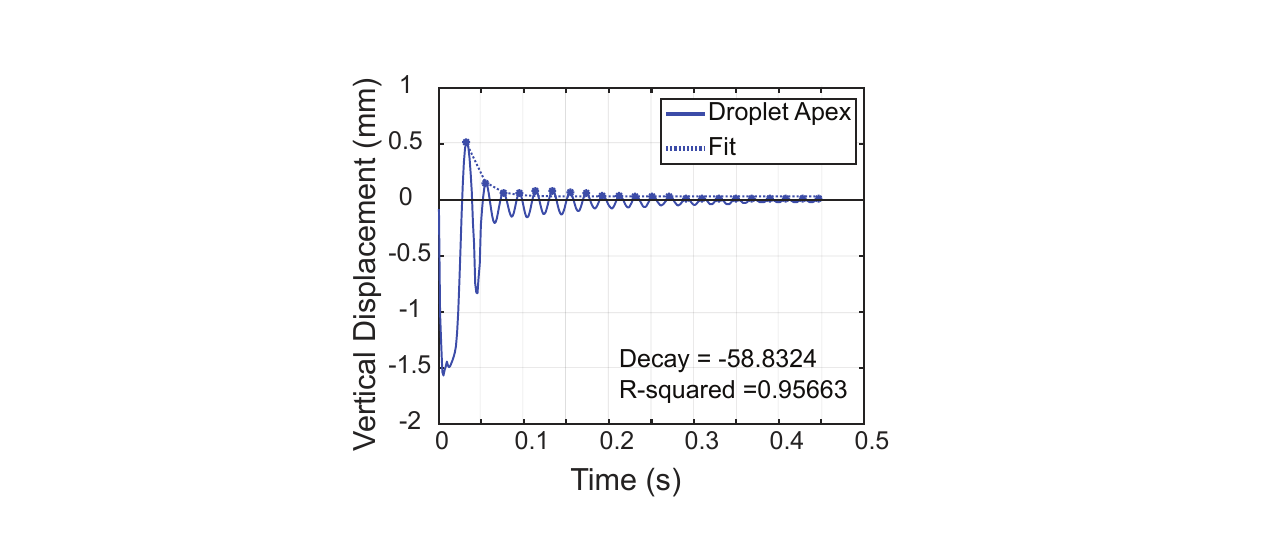} 
\label{}
\end{figure*}

\subsection{Spring and mass system}

\begin{figure*}[hbt!]
\centering
\includegraphics[width=1\textwidth]{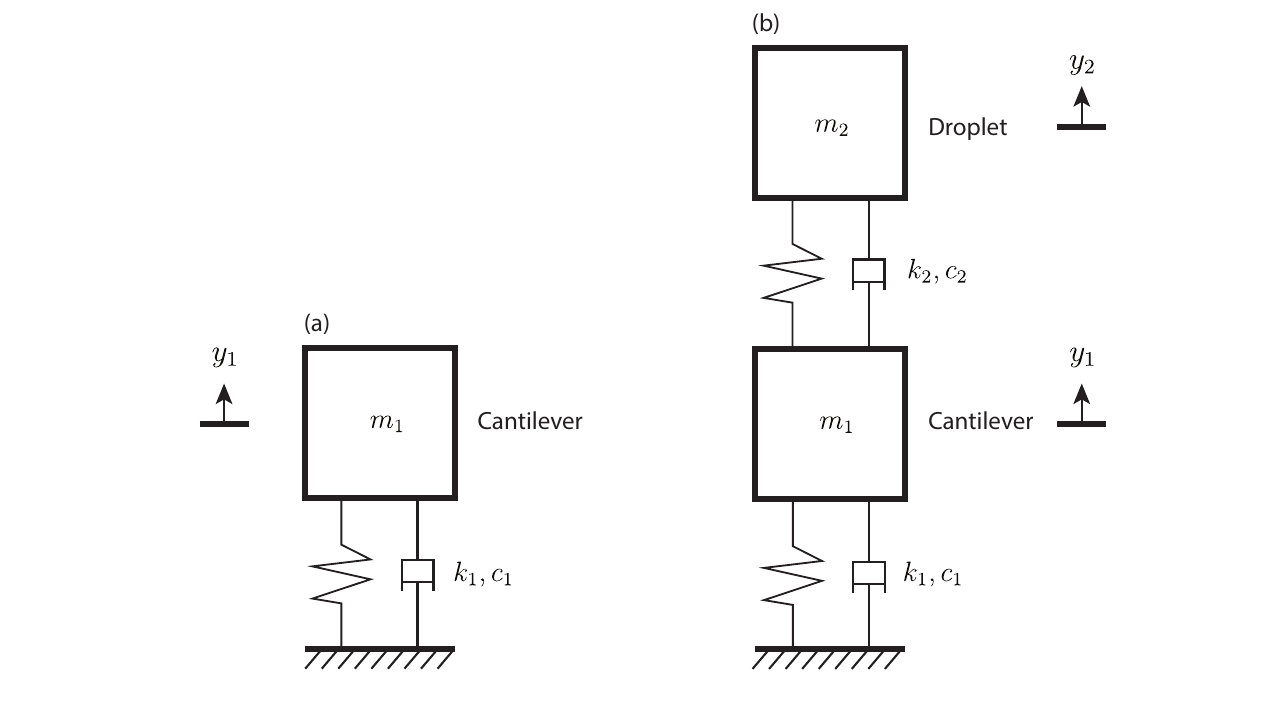} 
\label{}
\end{figure*}
Below is a visualization of the spring and mass system with a damper for the droplet-cantilever system. The derivation of a single mass and spring system are as follows (a): 
\begin{equation}
    m_{1}\ddot{y} + c_{1}\dot{y} +k_{1}y = 0,
    \label{single_system}
\end{equation}  
Rearranging the equation and then dividing by $m_{1}$, we get
\begin{equation}
    \ddot{y} = \frac{1}{m_{1}}[(-c_{1})\dot{y}-k_{1}y],
    \label{single_system}
\end{equation}  
Of which there are two solutions to the equation of motion. 
The derivation the cantilever in the droplet cantilever system is as follows (b):
\begin{equation}
    (m_{1}+m_{2})\ddot{y_{1}} + (c_{2}+c_{1})\dot{y_{1}} - c_{2}\dot{y_{2}} + (k_{2}+k_{1})y_{1} - k_{2}y_{2} = 0,
    \label{}
\end{equation}  
Rearranging and dividing by $m_{1} + m_{2}$, we get:
\begin{equation}
    \ddot{y_{1}} = \frac{1}{m_{1}+m_{2}}[-(c_{2} + c_{1})\dot{y_{1}} + c_{2}\dot{y_{2}} - (k_{2} + k{1})y_{1} + k_{2}y_{2}],
    \label{}
\end{equation}  

The derivation of the droplet in the droplet-cantilever system:
\begin{equation}
    m_{2}\ddot{y_{2}} - c_{2}\dot{y_{1}} + c{2}\dot{y_{2}} - k_{2}{y_{1}} + k_{2}y_{2} = 0,
    \label{}
\end{equation}  
Rearranging and dividing by $m_{2}$, we get:
\begin{equation}
    \ddot{y_{2}} = \frac{1}{m_{2}}[c_{2}\dot{y_{1}} - c_{2}\dot{y_{2}} + k_{2}{y_{1}} - k_{2}y_{2}],
    \label{}
\end{equation}  

MATLAB's ODE45 solver was used to solve for the displacement for the single cantilever system and the droplet-cantilever system. An example of the free vibration graph is shown in (a) and the droplet-cantilever system in (b) for a 40 mm length cantilever.
\begin{figure*}[hbt!]
\centering
\includegraphics[width=0.99\textwidth]{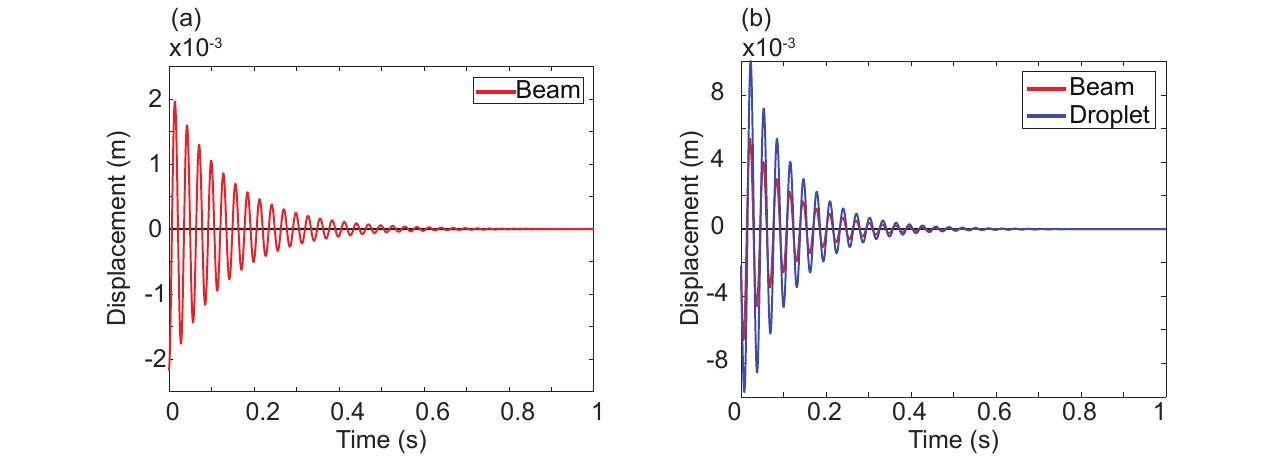} 
\label{}
\end{figure*}

\vspace{5mm}

\vspace{5mm}
\noindent\textbf{Competing interests}
The authors have no competing interests.

\noindent\textbf{Author Contributions}
SJ conceived the idea; CF, RM, MS, and SJ designed and built the experiment, and collected and analyzed data. CF and RM interpreted the results and wrote the manuscript. CF, RM, MS, and SJ revised the manuscript. 

\noindent\textbf{Data Accessibility}
Other datasets analysed during the current study are available from the corresponding author upon request.

\noindent\textbf{Funding}
Authors thank Mr. Yohan Sequeira for his initial contribution to this project. 
S.J. acknowledge funding support from the National Science Foundation (NSF) grant no. CMMI-2042740. 


\end{document}